\def\cm{{\rm\thinspace cm}}

\def\erg{{\rm\thinspace erg}}

\def\K{{\rm\thinspace K}}

\def\Msun{\hbox{$\rm\thinspace M_{\odot}$}}

\def\s{{\rm\thinspace s}}

\def\cmps{\hbox{$\cm\s^{-1}\,$}}

\def\pcmcu{\hbox{$\cm^{-3}\,$}}

\def\ergps{\hbox{$\erg\s^{-1}\,$}}

\def\spose#1{\hbox to 0pt{#1\hss}}
\def\approxlt{\mathrel{\spose{\lower 3pt\hbox{$\sim$}}
        \raise 2.0pt\hbox{$<$}}}
\def\approxgt{\mathrel{\spose{\lower 3pt\hbox{$\sim$}}
        \raise 2.0pt\hbox{$>$}}}

\documentstyle[psfig]{mn}

\title
[Coronal flares in Active Galactic Nuclei]
{Magnetic reconnection: flares and coronal heating in Active Galactic Nuclei}

\author[T.~Di Matteo]
{T.~Di Matteo\\
{Institute of Astronomy, Madingley Road, Cambridge, CB3 OHA}\\}

\begin{document}

\maketitle

\begin{abstract} 
A magnetically--structured accretion disk corona, generated by buoyancy
 instability in the disk, can account for observations of flare--like
 events in Active Galactic Nuclei.  We examine how Petschek magnetic
 reconnection, associated with MHD turbulence, can result in a
 violent release of energy and heat the magnetically closed regions of
 the corona up to canonical X-ray emitting temperatures.  X-ray
 magnetic flares, the after effect of the energy released in slow
 shocks, can account for the bulk of
 the X-ray luminosity from Seyfert galaxies and consistently explain
 the observed short--timescale variability.
\end{abstract}


\section{Introduction}

Active Galactic Nuclei (AGN) are believed to be powered by accretion.
UV and X-ray observations of AGN, particularly in Seyfert 1 galaxies,
indicate that the gravitational binding energy of a massive black-hole
is dissipated partly in a cold accretion disk and partly in a hot,
tenuous corona above it.  Thermal Comptonization of soft UV-radiation
in the corona leads to the production of a hard X-ray continuum, some
of which is reprocessed by the cold, dense disk (Haardt \& Maraschi
1991, 1993) giving rise to the observed reflexion features in the
spectra.  Recent work, both on the observational and theoretical side, 
(Sincell \& Krolik 1997, Nandra 1997, Stern et al. 1995) indicates
that the UV continuum in AGN may be mostly produced by reradiation of
energy absorbed from X-rays irradiating the accretion disk. Also, in
order to explain the different ratios of X-ray and UV luminosities in
different objects it has been suggested that the corona consists of
localized active regions (e.g. Haardt, Maraschi
\& Ghisellini 1994).  
Popular models for the production of the X-rays, therefore, suppose
that a large part of the disk's dissipation takes place in a small
amount of mass in the hot corona.  Variability timescales of the order
of few hours observed in AGN also put upper limits on the size of the
emitting regions, and imply that enormous amounts of energy need to be
released on very short timescales. It is therefore important to
understand what processes can efficiently channel significant amounts
of energy into the hot coronal medium and the way in which energy can be
released in localized flare--type events.

In a differentially rotating disk the evolution of a magnetically
structured corona is very hard to suppress.  The strong magnetic
fields, continuously generated by the dynamo action in an accretion
disk, are strongly buoyant and are forced to invade the region
sandwiching the disk itself. Once outside the disk, the magnetic flux
tubes can reconnect efficiently and dissipate part of the accretion
energy in localized active flares. Buoyancy, therefore, constitutes a
mechanism that channels part of the energy released in the accretion
process directly into the corona outside the disk. Magnetic
reconnection can be responsible for the rapid dissipation of
magnetic energy though a field-aligned electric potential in thin
current sheets.

We, here, describe how the observationally required large heating
rates per unit mass, in the optically thin gas of a hot coronal
region, can be accounted for by an ion--acoustic instability in the
context of slow shocks associated with Petschek type reconnection and
give rise to flare--like events. We estimate that such a process may
be responsible for the heating of the coronal plasma in the active
regions of an AGN to a level where it emits X-rays at a temperature of
$\sim 10^9 \K$ even in the presence of Inverse Compton and synchrotron
cooling processes.  Magnetic buoyancy, which drives magnetic flux out
of the disk, naturally decreases the plasma density to the point where
the active region can reconnect very efficiently. X-ray observations
imply that the coronal plasma is very tenuous with a density much
lower than that of the underlying disk.  We show that the energetics
of such flares are consistent with the X-ray luminosities of typical
Seyfert galaxies. The onset of magnetic flares through slow shocks
associated with Petschek reconnection, should occur over timescales
comparable to those required to explain X-ray variability observed in
AGN.

This picture is also, by analogy, validated by X-ray observations of
the solar corona.  Recent {\it Yohkoh} observations (Tsuneta 1996,
Yokoyama \& Shibata 1995 and references therein) of solar flares show
clear evidence for magnetic reconnection taking place in
magnetically confined loop-like volumes. There is evidence that
reconnection serves as a highly efficient engine for converting
magnetic energy into kinetic and thermal energy with standing slow
shocks.  Moreover, it is now well established that in the Sun active
regions are formed from the emergence of magnetic flux. It has been
shown (Shibata et al. 1989) that magnetic reconnection can be driven
self-consistently by the magnetic buoyancy instability (the Parker
instability) between emerging flux and the overlying coronal magnetic
field.

\section{Emerging magnetic flux and reconnection in AGN coronae}
Magnetic fields have long been considered an important element in the
dynamics of accretion disks in particular as a mechanism for supplying
the internal stresses required for efficient angular momentum
transport (Coroniti 1979, Stella \& Rosner 1984, Balbus \& Hawley
1991).  Strong fields in the accretion disk are continuously generated
because the shearing and the turbulent disk flow serve as an
effective dynamo that can rapidly strengthen any seed field. Field
amplification is limited by non--linear effects: as a consequence of
buoyancy (Parker instability) strong fields will rapidly emerge from
the disk leading to an accretion disk corona consisting of many
magnetic loops (Galeev Rosner \& Vaiana 1979). Part of the accretion
energy is therefore released into a corona in the form of magnetic
energy.  Either by internal shear or by encountering already present
magnetic flux tubes, reconnection takes place and rapid magnetic
energy dissipation can occur through the formation of a slow shock in
a thin current sheet.

Not all of the emerging flux will produce a flare. Simple loop flares
are triggered only if sufficient flux emerges so that the current
sheet occupying the interface between new and old flux reaches a
critical height that depends on the emergence speed and magnetic field
strength (e.g. Heyvaerts, Priest \& Rust 1977 and references therein).
As the sheet rises through the corona its temperature increases until
it achieves a state where Joule heating is balanced by thermal
conduction.  At this point, the current sheet can reach the critical
temperature for the onset of turbulence (creation of a slow shock).
This is the main phase of the flare, the current sheet reaches a
steady state, with reconnection based on a marginally turbulent
resistivity (anomalous electric resistivity), with X-ray flares being
the after effect of the energy release by the slow shocks.

In summary, in a differentially rotating
disk gas, the Parker instability should lead to the formation of an
electrodynamically coupled corona. The onset
of impulsive reconnection corresponds to the evolution of anomalous
dissipation in localized regions, in a similar manner to flare events in the
Sun's corona.  Although the plasma regime envisaged here for the
accretion disk corona is vastly different from that of the Sun, its
structure is quite similar. In both cases the
low-$\beta$ coronal plasma is dynamically controlled by magnetic
fields the footpoints of which are embedded in a buoyantly unstable,
$\beta \sim 1$ plasma.

It is now well established that active regions in the Sun are formed by
the emergence of magnetic flux tubes. The soft X-ray telescope aboard
of {\it Yohkoh} has provided the opportunity to observe the
interaction of emerging flux with overlying coronal magnetic fields
with high spatial and temporal resolution (e.g. Tsuneta 1996). In addition,
numerical simulations of magnetic reconnection, driven by magnetic
buoyancy instability, have been performed by solving resistive
magnetohydrodynamic equations assuming uniform magnetic field at the base
of the Sun's corona. The emergence of flux sheets is reproduced by
simulating the Parker instability. As the instability develops, gas
slides down the expanding loop, and the evacuated loop rises as a
result of the enhanced magnetic buoyancy (expansion is due to magnetic
pressure force although the dynamics are controlled by the downflow,
due to gravity, along the rising loop). Consequently, as long as the
magnetic pressure at the top of the magnetic loop is larger than the
coronal (gas and magnetic) pressure, the expansion of the magnetic
loop continues. In particular detailed simulations of such instability
(Shibata et al. 1989) have shown that during the expansion of the
magnetic loop the rise velocity and the local Alfv\'en speed of the
loop increases linearly with the height $h$ above the disk whereas the
density in the sheet scales as $\rho
\propto h^{-4}$ (the normalizing constant is $H$, the scale height of the
basis where the plasma $\beta$ distribution initiates the instability).

In order to meet the observational constraints, the coronal plasma in
an AGN needs to be very tenuous (optically thin). The optical depth
deduced from X-ray observations is $\tau\approx 1$
(e.g. Poutanen 1997 -- where $\tau$ is the Thomson optical depth given
by $\tau=n
\sigma_{\rm T} R$, $n$ is the electron number density 
and $\sigma_{\rm T}$ the Thomson cross section).  Since AGN show variations
in luminosity on very short time scales, $R$, the thickness
of the emitting region, can be constrained from observations of the
shortest variability timescales ($R/c \sim 200\s$) to be of the order $R
\sim 6
\times 10^{12} \cm$. For $\tau$ in the given range,
 $n \sim 3 \times 10^{11} \pcmcu$.  If the energy released in a flare
needs to account for an impulsive event and provide the required
energy dissipation rates, reconnection needs to be very
efficient. Clearly, as the reconnection velocity increases ($\propto
V_{\rm A}$ the Alfv\'en speed; $V_{\rm A}=B/\sqrt(4 \pi n m_{\rm p})$
where $B$ is the magnetic field strength and $m_{\rm p}$ the proton
mass) the timescale for reconnection decreases and the energy
dissipation rate increases. Therefore, according to magnetohydrodynamic
theory, reconnection is most efficient when the current sheet density
is low or equivalently when the loop has been pushed up high enough in
the corona. As mentioned above, the density would naturally decrease as
the magnetic tube expands and rises from an underlying dense disk
(with, typically, $n_{\rm disk} \sim 10^{16}
\pcmcu, T_{\rm disk}\sim 10^5 \K$) due to the gravitational downflow
caused by the magnetic buoyancy instability and reach values of the order
of $n$ as determined above form observational constraints.  
If we assume that a flare is triggered when the density has decreased to
value $n_{in} \sim 2\times 10^{12} \pcmcu$ ($n_{in}$ is the density of
the material entering the shock: a factor $5-6$ greater than $n$ as
obtained from the jump condition, eqn. (6)) we can estimate the
critical height at which reconnection could suddenly start in an AGN
corona.  According to the afore described magnetic buoyancy
simulations, the loop expansion is self-similar and the loop top
reaches a height given by
\begin{equation}
h_{\rm flare} \sim ({10^{16}}/{2 \times 10^{12}})^{0.25} H \sim 8 H ,
\end{equation}
where $H$, in this case, is the pressure scale height of the accretion
disk (in a standard Shakura \& Sunyaev model $H/R \approxlt 0.1$, so
that in the central regions ($R=3-4 R_{\rm S}$), $h_{\rm flare}$ would
correspond to about $3 R_{\rm S}$, where $R_{\rm S}$ is the
Schwarzchild radius). This implies that a magnetically structured
corona can exist at a few scale heights above the disk.  This is
consistent with theoretical arguments related to buoyancy: the greater
shear energy in the larger magnetic flux tubes in the disk amplifies
the magnetic field beyond equipartition and therefore allows only
the large scale length flux ($R \approxgt H$) to be expelled from the
disk. (There are no obvious constraints from the spectra of Seyfert
galaxies that can rule out the occurrence of flares at scale heights larger
than $H$).

Given the relevance of emergence flux models in coronae, it would be
appropriate to treat magnetic reconnection in a self-consistent manner
as the interaction and coalescence of the rising flux tubes.
 Such a treatment is still a very difficult task as it
would require a run of densities magnetic field with $h$. 
We will therefore consider only the current sheet
in a self-consistent manner adopting values for the density
interpreted from observations of AGN X-ray emission.


\section{Magnetic reconnection and coronal heating}
In this section we describe the dynamics of the formation of localized
field--aligned electric fields in the corona of an AGN within the
context of a simple analytical model.  

The major problem faced by any
mechanism which attempts to dissipate the stored magnetic energy is
the low resistivity of the pre--flare coronal plasma, with associated long
magnetic diffusion times and high magnetic Reynolds numbers
($R_{m}=\tau_{\rm D}/\tau_{\rm A}$, where $\tau_{\rm D}$ is the
diffusion timescale and $\tau_{\rm A}$ is the Alfv\'en time). Since
the diffusion timescales varies as the squares of the characteristic
length scale, $l$, associated with changes of magnetic field, it is 
desirable to make $l$ as small as possible.  The topology of
reconnection offers a way of achieving this, i.e. two sets of
oppositely directed fields in close proximity -- a so called neutral
sheet configuration.

The current density in a neutral sheet is given by
\begin{equation}
j=nev=\frac{c}{4\pi}(\nabla \times {\bf B}) \sim \frac{cB_{x}}{4\pi l}.
\end{equation}
As $l$ becomes small, $j$ becomes large and the Ohmic dissipation
$\eta j^2$ can become very large, where $\eta$ is the magnetic
resistivity.  Moreover, where magnetic reconnection is required to
explain various explosive phenomena, Petschek models need to be
invoked. Petschek reconnection takes into account the inclusion of the
effect of waves -- the occurrence of a slow mode MHD shock -- and is
almost independent of Reynolds numbers.  The development of plasma
instabilities (turbulence in the shock) implies the onset of anomalous
resistivity much larger than the classical resistivity. The electrons
velocities are randomized by scattering on the waves and, in this
manner, the dissipated energy goes mainly into heating of the resonant
electrons. A well--studied example of such a mechanism (and probably
most important mechanism in the neutral sheet) is the ion acoustic
instability that sets in when the electron drift velocity surpasses
the ion-acoustic wave velocity (e.g. Coroniti \& Eviatar 1977).  This
provides fast reconnection rates leading to explosive
outbursts. Simulations (Yokoyama \& Shibata 1994) have shown that for
magnetic reconnection driven by the Parker instability anomalous
instability indeed leads to (fast) Petschek--type reconnection.

In the following section we describe the magnetic structure and dynamics
of the reconnection region of the flare. We employ the theory of
Petscheck reconnection (X--point with slow shock
structure) which appears to be consistent with the detailed
observations of the Sun's coronal flares.


\subsection{Slow standing shock}
We here review the basic set of equations which are useful for our
calculations.  The inflow velocity toward the X-point and the slow
shocks is $V$, and the outflows have the  Alfv\'en speed in both the
Sweet--Parker or in the Petschek theories (momentum balance). The
latter theory also predicts that the acute half angle of the slow
shock, $\alpha \approx 1-2^o$ (this is supported by observations). Mass
conservation then gives
\begin{equation}
n_{\rm in}V L \cos\alpha = n_{\rm out}V_{\rm A}L \sin\alpha,
\end{equation}
where $L$ is the length of the region of close, oppositely directed
magnetic field lines (the length of the slow shock region). The
suffixes $n_{\rm in}$ and $n_{\rm out}$ indicate the inflow and the
outflow respectively. We thus obtain
\begin{equation}
V=V_{\rm A} \tan\alpha \frac{n_{\rm out}}{n_{\rm in}}.
\end{equation}
We next examine the standing slow shock attached to the X--point.  It is
usually assumed that the slow shock is switch--off, and that the
inflow speed $V \approxlt V_{\rm A}$. The momentum flux across the shock
must be continuous giving
\begin{equation}
p_{\rm in} + \frac{B_{\rm in}^2}{8\pi }=p_{\rm out},
\end{equation}
where $p_{\rm in}$ and $p_{\rm out}$ are the inflow and outflow gas
pressures respectively. Since the thermal conduction along the
reconnected field lines is high, we expect the slow shock to be
isothermal within the thermal conduction front (across the shock region).
Therefore the jump condition for the plasma density is given from
eqn. (5)
\begin{equation}
\frac{n_{\rm in}}{n_{\rm out}}=1 + \frac{1}{\beta_{\rm in}},
\end{equation}
where the plasma $\beta$ is given by $\beta=p_{\rm in}/(B_{\rm in}^2/8
\pi)$, which for Petschek reconnection is typically small (of the order
of $0.2-0.3$, giving a density jump of $\sim 5-6$).

In order to set the typical magnetic field strength, $B$, in the rising loop
we use the condition for the onset of magnetic buoyancy in the disk.
If the shear stresses acting on a flux cell generate
a magnetic pressure which exceeds the gas pressure, upward buoyancy
forces will rapidly remove the flux tubes from the disk (Rosner \&
Vaiana 1979 and Coroniti 1981). The field strength in the loop is set by
the condition that $\beta \sim 1$ or
\begin{equation}
\frac{B^2}{8\pi} \approx n_{\rm dis} k T_{\rm disk}.
\end{equation} 
In order to maintain pressure balance, magnetic flux tubes with such strong
magnetic fields will contain less plasma than their ambient
surroundings; therefore, they
 become subject to buoyancy and will emerge from the disk as
described in the previous sections. For a typical AGN, the disk temperature
is of the order of $T_{\rm disk}\sim 10^5\K$ and its number density $
n_{\rm disk} \sim 10^{16} \pcmcu$. $B$ is therefore $\sim 2 \times 10^3$
Gauss. With this value of $B$, $V_{\rm A}= B/\sqrt{(4 \pi n_{\rm out}
m_{\rm p})} \sim 10^9
\cmps$ where $n_{\rm out} =n\sim 3 \times 10^{11}\pcmcu$ , $V=1/10V_{\rm A}$ and from
 continuity, $l/L \approx V/V_{\rm
A} = 1/10$.  We take the dimension $l$ to be set by shortest
variability timescale i.e. $l=R=6\times 10^{12}\sim 2 R_{\rm s}$ where
$R_{\rm s}$ is the Schwarzchild radius for a mass $M=10^7 \Msun$,
typical of a Seyfert.

\subsection{Energetics}
We next examine the energetics of a reconnection site when the main
phase of a flare is triggered i.e. when turbulence is created with an
associated MHD slow shock. Our aim is to obtain order of magnitude
estimates of the different processes which contribute to the heating
and cooling of the current sheet and to derive an estimate of the energy
released in a flaring corona in an AGN.

The plasmas on the reconnected filed lines are significantly heated by
the two standing slow shocks.  In addition to the heating,
radiative losses, especially due to Inverse compton and synchrotron
processes, become important in an AGN environment and will also be taken
into account.
  

We consider one single sheared coronal loop, i.e. a magnetic flux tube
with a current flowing due to the magnetic shear.  We assume that the
energy produced in the reconnection site goes into slow MHD shock
heating as well as energy of the bulk flow. This assumption is good
for the Petschek reconnection model, where the slow shock angle is
small and the outflow plasma $\beta_{\rm out}$ is high.  Energy
balance therefore gives
\begin{equation}
\frac{B_{\rm in}^2}{4 \pi} V L= \frac{1}{2} (m_{\rm p} n_{\rm out} V_{\rm A}^2 + 5 n_{\rm out} k T)V_{\rm A} l + \kappa_{0} \frac{T^{7/2} }{L} l + E_{\rm r}.
\end{equation}
i.e. the energy dissipation rate has to balance with the inflow
magnetic energy.  The term on the left hand side of equation (8)
represents the magnetic enthalpy flux which coincides with (1/4) of
Joule heating of the current sheet ($\eta j^2 4 L l$) (eqn. 2).  The
first two terms on the right hand side are the kinetic energy flux and
the enthalpy flux of the outflow. The third term on the right hand
side is the heat flux and represents an estimate of the heat produced
in the slow MHD shock from conduction losses.  The term $\kappa_0
T^{5/2}(T/L)$ is the Spitzer thermal conductivity ($\kappa_0 = 1.79\times
10^{-5}/ \ln\Lambda$) and $T$ is the temperature of the slow MHD
shock.  The forth term represents the relevant radiative energy
losses. The electrons in the coronal loop are cooled though inverse
Compton scattering and synchrotron radiation. Hence
\begin{eqnarray}
E_{\rm r}/ 4Ll&= &q_{\rm Comp} +q_{\rm synch} \nonumber \\
&=&4\left( \frac{{\rm k}T}{m_{\rm e} c^{2}}\right) \sigma_{\rm T} n_{\rm out} cU_{\rm rad} +
\frac{2\pi}{3}\frac{k T}{c^{2}} \frac{\nu_{c}^{3}}{l},
\end{eqnarray}
where $\sigma_{\rm T}$ is the Thomson cross section and $U_{\rm rad}$
is the energy density of the isotropic soft photon bath. The seed
photons are partly created locally by thermal synchrotron processes
and partly created in the underlying disk. The photons from the disk
are in turn due to both local energy dissipation and to re--radiation
of hard X-rays created in the active coronal loops.  Recent work
(Nandra 1996; Sincell \& Krolik 1997) indicates that most of the
quasi-thermal emission from the disk is dominated by reradiation of
the energy absorbed from the X-rays irradiating the disk.  This
implies that $U_{\rm rad}$ can not exceed $ U_{\rm mag} = B^2/8\pi$
(i.e. the total energy density stored in the magnetic structure) and
depends on how efficiently magnetic field is dissipated.  A reasonable
photon energy density is then given by $U_{\rm rad} =(B^2/8\pi) V_{\rm
A}/c$ where $c$ is the light speed (see also Di Matteo, Blackman \&
Fabian 1997).  $\nu_{\rm c}$ is the self--absorbed synchrotron
frequency is obtained by equating the Rayleigh-Jeans emission to the
synchrotron emission (Zdziarski 1985).

The classical thermal conductivity, $E_{\rm H}=\kappa_{0} T^{7/2}/ L $
only represents a lower limit to the heating that goes into the slow
shock region. When the anomalous resistivity (with a slow MHD shock)
develops the conductivity reduced (i.e. it saturates) with respect to
its classical value. Taking plasma turbulence to be the source of
energy release in the flare, the field--aligned thermal flux becomes
anomalous and can be approximated by (Manheimer 1977; see also Somov
1992)
\begin{equation}
F_{\rm an}=\frac{n_{\rm in}(k T)^{3/2}}{4 m_{\rm e}^{1/2}} ,
\end{equation}
where the simple analytical function $F_{\rm an}$ is obtained under the
assumption that a current sheet can reach an equilibrium state
specified by  saturated ion-acoustic turbulence.
\begin{figure}
\centerline{\psfig{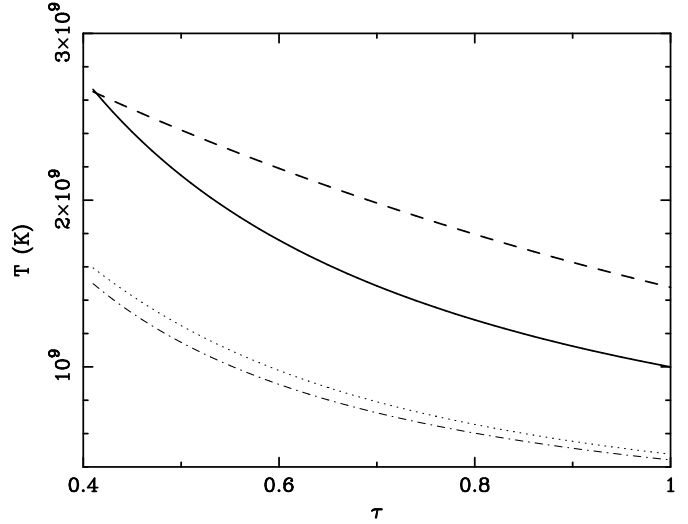}}
\caption{The temperature solution for a current sheet producing a flare
in an AGN for different values of the optical depth $\tau$.  The
dashed and solid lines represent the temperature solution for $U_{\rm
rad} =(B^2/8\pi) V_{\rm A}/c$.  The thinner lines (dotted and
dot-dashed) show the effect of considering greater Inverse Compton
losses, i.e. $U_{\rm rad}({\rm max})=(B^2/8\pi)$.  The solid and
dot-dashed lines take into account the effects of the slow MHD shock
on the conduction term (anomalous conduction $F_{\rm an}$ as opposed
to $E_{\rm H}$).}
\end{figure}
We can now solve equation (8) for both $F_{\rm an}$ and $E_{\rm H}$.
With $B$ and all the relations between velocities and densities for
the outflow and the inflow derived as in the previous section, we can
solve for the temperature of a flare in an AGN corona, for a range of
$\tau$ (densities). The resulting temperature solutions with the two
respective conduction terms is shown in Fig 1. We show temperature
solutions for $U_{\rm rad} = (B^2/8\pi) V_{\rm A}/c $ and for the
extreme value, $U_{\rm rad}({\rm max}) = B^2/8\pi$ (Inverse Compton
losses usually dominate synchrotron ones).  This clearly illustrates
that temperatures of the order of a few $\times 10^9 \K$ can be easily
achieved, even in the presence of intense radiation fields, implying
that magnetic reconnection can indeed provide for the heating of the
plasma to a level where it emits X-rays. Observations of spectral
cut-offs in AGN directly determine the temperature of the X-ray
emitting gas to be of this order (e.g. see Poutanen et al. 1997).
It is worth stressing that because of the strong dependence of the
shock heating term on $T$, the temperature solution, $T\sim 10^9\K$ is
very stable and only weakly dependent on changes in current sheet
dimensions or $B$. 

Note that for $\tau > 1$ (and corresponding density) the conduction
cooling term acts very efficiently and the temperature of the flare
becomes $< 10^9 \K$. Flares with such $\tau$ would only contribute to
emission in soft X-ray band and, given the strong dependence on $T$ of the
conduction flux, they would be energetically negligible. This is in
accord with interpretation of X-ray observations which imply $\tau
\approxlt 1$, (also discussed in the context of coronal magnetic
flares by Nayakshin \& Melia 1997).

The magnetic energy is converted into plasma heating
in the slow shock regions to give the temperature just
determined. It is important to check whether the total energy
production rate for flares, as estimated from the amount of energy
going into the shock (i.e. to the conduction flux), is
consistent with the energy output of a typical AGN. For $T \sim
2-3\times 10^9 \K$, the energy of one flare is then given by
\begin{equation}
E_{\rm flare}= F_{\rm an} (4 Ll) \sim 5 \times 10^{41} \ergps.
\end{equation}
This implies that we only require $N \sim 10-20$ flares to be active
at any given time in order to account for typical Seyferts luminosity
-- of the order of $10^{42-43}\ergps$ (actually Eqn. (11) only gives a
lower limit to the total energy going into the shock and into partile
heating: energy can also be tapped from internal and kinetic energy
terms).  Therefore, the energy produced in slow MHD shocks is
completely consistent with the X-ray energy output of an AGN.

We have shown, by means of order of magnitude estimates, that the slow
shock region in the current sheet of an AGN corona is
heating-dominated and can give rise to a flare.  In magnetically
structured coronae in AGN or Galactic black hole candidates, the
hard X-rays are produced by Inverse Compton scattering on $10^9 \K$
electrons, so clearly the energy flow into the electrons has, in
some way, to be guaranteed.  Assuming that the current sheet becomes
turbulent implies that dissipation will take place in a spectrum of
length--scales, from a dominant scale down to the smallest scales, and,
therefore down to the shortest possible timescales.  On average, the
timescale associated with the flaring of such a magnetic loop is, in
accordance with Petsheck models completely consistent with the typical
variability timescales of Seyferts. Petschek (1964) in fact
proposed that the inclusion of the effect of waves enables
reconnection to be as short as a few Alfv\'en times; $\tau_{\rm A}=
l/V_{\rm A} \sim 100 $ a few thousand seconds depending on $B$ and
$l$ and almost independent of the Reynolds number.

\section{Summary and Discussion}
We have presented a simple model that takes into account the
impulsive dissipation of magnetic energy during coronal flares in AGN.
According to standard models, the X-rays are produced by Inverse
Compton scattering of lower energy photons on energetic
electrons. We show that the energy flow to the electrons that is
needed for this process is guaranteed by the release of coronal
magnetic energy. The onset of an ion-acoustic
instability associated with slow MHD shocks and Petschek reconnection,
heats the flare plasma to X-ray temperatures $\sim 10^9
\K$, as required by observations.  
The luminosity produced in a
magnetically structured, flaring corona is completely consistent with
typical power outputs of AGN if at least $N \sim 10$ flares are
triggered at any given time.  The plasma reaches low enough densities,
(high Alfv\'en speeds), when reconnection is driven by the Parker
instability, such that flares are typically triggered at $h \sim 8 H$
above the accretion disk (coronal structures can have scale-heights
greater than the disk).  
Energetic flare--type events
naturally explain the observations of short timescale variability in
AGN. 

The buoyantly unstable magnetic flux tubes,
once outside the disk, rise through the coronal atmosphere at their
local Alfv\'en speed. The timescale for them to rise up to a few scale
heights therefore is much shorter than the disk dynamical
timescale (and less than the shear timescale in the disk $\sim
2/3\Omega$). This implies that tubes will not be disrupted even
when flares are triggered a few scale-heights above the disk.

An important problem faced by any model of AGN coronae, is to
determine the population of non-thermal particles produced by the slow
MHD shock and the effects of the impulsive electric field.  According
to theoretical arguments for the kind of electric fields of interest,
the turbulence is characterized by an highly anisotropic distribution
of ion-acoustic waves.  In these circumstances it has been shown
(Heyvaerts et al. 1977) that electron heating occurs with practically
no acceleration. Because electrons travel much faster than
ion-acoustic waves, the resonance occurs mainly with waves propagating
normal to the particle velocity: this leads to angular diffusion with
little change in particle energy. This implies that electrons can be
characterized by a mean increase of energy or 'temperature'.  In this
sense we can treat the hot current sheet as a 'thermal' source of
electrons as required by the observations of spectral cut-offs in the
X-ray spectra of AGNs.  Note, though, that the presence of an electric
field and plasma turbulence in the sheet would inevitably cause
acceleration of charged particles to a certain degree. This is
particularly relevant in the case of Galactic black hole candidates
where recent high energy X-ray observations require the presence of
non-thermal electron tails (e.g. Poutanen \& Coppi 1988). Further
investigations are needed but they are beyond the scope of this paper.

The geometry of reconnection described here does not
provide the only viable way for the process to occur.  
In an accretion disk corona, the closed magnetic
loops will get twisted by the rotation of the accretion disk. As the
twist accumulates, the magnetic loops expand and finally approach the
open field configuration. A current sheet is formed inside the
expanding loops and, in the presence of resistivity, magnetic
reconnection will take place. Open field lines anchored to the disk,
may be the region where winds or jets blow from the disk;
 these could play an important role in the dynamics
of the accretion disk itself. The luminosity of flares will then vary
with respect to the different values of $B$ in different morphologies in which
reconnection takes place, and with the size and the total number of
flares.  All of these factors can give rise different covering
fractions of the X-ray emitting regions which can result in very
different variability timescales as observed in AGN.  Finally,
magnetic buoyancy is not strictly necessary in order to maintain a
certain level of activity. Once a flare has been triggered,
reconnection can be maintained by positive feedback (the fast
outflow due to reconnection rarefies the reconnection region, thinning
the current channel in this way maintaining the anomalous resistivity
at the neutral sheet).

\section*{Acknowledgements}
I acknowledge PPARC and Trinity College  for financial
support.  I thank my supervisor A.C Fabian, E. Blackman, M.J. Rees
and in particular Jean Heyvaerts for discussions.

\end{document}